\documentclass[12pt]{article}
\usepackage{amssymb}

\usepackage{amsmath}
\usepackage{color}
\usepackage{graphics,latexsym,amsfonts}
\usepackage{graphicx}
\usepackage{pstricks, pst-node}
\usepackage[round]{natbib}
\usepackage{subfigure}
\usepackage{eurosym}
\newcommand{\rio}{\mathbb{R}}

\begin{document}

\thispagestyle{empty}

\begin{center}
\large

\textbf{The Financing of Innovative SMEs: a multicriteria credit rating model }\\[1pt]
\medskip
SILVIA ANGILELLA\footnote{%
Email address: \texttt{angisil@unict.it}} \hspace{1cm} SEBASTIANO MAZZ\`{U}\footnote{%
Email address: \texttt{s.mazzu@unict.it}}
\medskip
\\[0pt]
\footnotesize{Department of Economics and Business}\\
\footnotesize{University of Catania},\\
\footnotesize{Corso Italia, 55, Catania 95129, Italy}\\

\today
\end{center}

%
\begin{abstract}
Small Medium-sized Enterprises (SMEs) face many obstacles when they try to access  credit market. 
These obstacles are increased if the SMEs are innovative. 
In this case, financial data are insufficient or even  not reliable. Thus, when building a  judgemental rating model, mainly based on qualitative criteria (soft information),  it is very important to finance SMEs' activities. Until now, there isn't a multicriteria credit risk model based on soft information for innovative SMEs. In this paper, we try to fill this gap by  presenting a multicriteria credit risk model, specifically, ELECTRE-TRI. 
To obtain robust SMEs' assignments   to the risk classes,  a SMAA-TRI analysis is also implemented. In fact, SMAA-TRI incorporates ELECTRE-TRI by considering different sets of preference parameters and uncertainty in the data via Monte Carlo simulations.

Finally, we carry out a real case study, with the aim of illustrating the multicriteria credit risk model proposed.



\end{abstract}



\section{Introduction}

Since  small and medium-sized  enterprises (SMEs) are the backbone of all economies, 
scholarly attention to their credit risk assessment has considerably increased. 

SMEs face many financial restrictions  to access the credit market (for example see \citealp{Berger:2005}, \citealp{Beck:2006} and \citealp{Canales2012}).
Moreover, if the SMEs are  newly created firms innovative, these restrictions are more tight (see  \citealp{Brown:2009}).

Consequently, there exists a finance gap within the SME's sector, and such a condition is worsened when SMEs  have to request credit to the banks. Banks usually adopt different lending approaches that can be grouped in four categories: financial statement lending (based on the evaluation of information from balance-sheet data), asset-based lending (based on the provision of a collateral), credit scoring models (based on hard information), and relationship lending (see \citealp{Moro:2013}). 

More precisely, the credit scoring models are grouped in the families:
statistical, operational research and intelligent  techniques (see \citealp{Ravi:2007} and \citealp{Crook:2007} for a review on the topic). In financing SMEs, there is a recent research debate on which is the best type of lending technique.

From one hand, there are several papers supporting the idea that banks prefer to finance SMEs on the basis of  a long and strong  activity of relationship banking,   a type of lending mainly based on non financial information (soft information) (see among others  \citealp{Moro:2013} and \citealp{Berger:1996}).

On the other, there is a recent stream of research that asserts that big banks lend more easily to SMEs if lending techniques based on hard information are adopted (see for example \citealp{Berger:2006} and \citealp{Beck:2011}). 

Owing to the SMEs' lack of sufficient or reliable track records,  a qualitative rating based on the judgment of some experts (judgemental rating)  seems  the most useful approach to evaluate the SME's creditworthiness.

In fact, the significant role of the non-financial criteria for the SME's credit risk  evaluation is  supported by several research papers (see  among others \citealp{Altman:2010}, \citealp{Hernadez},  \citealp{Berger:2005} and  \citealp{Grunert:2005}; and specifically, for innovative SMEs see for example \citealp{Shefer:2005} and  \citealp{Hottenrott:2011}). 


Until now, to the best of our knowledge, there isn't a judgemental credit scoring model for innovative SMEs. In this paper, we try to fill this gap by  adopting a  \textsf{M}ultiple \textsf{C}riteria \textsf{D}ecision \textsf{A}id (\textsf{MCDA})  approach (\citealp{Vincke:1992}).
Since the financing of innovative SMEs is an uncertain  problem due to several and conflicting aspects, the Multiple Criteria Decision Aid approach (also called the constructive approach \citealp{Roy:1993}) seems the most useful (see  \citealp{Greco:2013} for some useful comments on this topic).

Several multicriteria approaches have already been adopted to predict business failures, that is a typical sorting problem (see \citealp{Zopounidis:2002}).
Most of them rely on a utility function to  classify enterprises into two categories: the default and non-defaulted (see among these  the multicriteria hierarchical discrimination approach proposed in \citealp{Doumpos:2002}).
Even some multicriteria approaches relying  on a utility function have been proposed with the aim of  helping credit granting decisions; for example the multicriteria methodology MACBETH (introduced in \citealp{Bana:1997} and \citealp{Bana:1999}) has been implemented as a qualitative credit rating model   in the banking sector (see \citealp{Bana:2002}).



Very few recent papers have also developed some credit rating models based on an outranking relation; for example,
ELECTRE TRI (firstly, introduced in \citealp{Yu:1992})  has been implemented for the first time as a quantitative credit rating model  in  \citealp{Zopounidis:2011} and  PROMETHEE II (\citealp{BransVincke85})   applied  for banks' rating evaluation (\citealp{Zopounidis:2010}). Moreover, a first attempt to estimate SMEs' performance, based on a financial ratio analysis, can be found in \citealp{voulgaris:2000}, where the multicriteria sorting approach, UTADIS (\citealp{Lagreze:1995}) has been applied. Recently, in \citealp{Nemery:2012} the innovation performances of SMEs have been assessed by the Flow-Sorting method. However, none of these papers attempts  to develop a judgemental rating model to assign innovative SMEs into risk classes. 
The aim of the  paper is to fill this gap by  presenting a multicriteria model to assist financial institutions in the evaluation of the risk involved in financing innovative projects.

First, we propose  ELECTRE TRI to evaluate the SMEs' credit risk,  mainly on the basis of qualitative criteria. 
To obtain robust SMEs' assignments   to the risk classes,  a SMAA-TRI analysis (\citealp{Tervonen2007}) has also been  implemented. In fact, SMAA-TRI incorporates ELECTRE-TRI by considering different sets of preference parameters and uncertainty in the data with  Monte Carlo simulations.

Then, to illustrate the multicriteria methodology proposed a real case study on four  Italian innovative enterprises in the phase of start-up is considered. The whole multicriteria approach has been implemented by interviewing some loan officers  from one of the main  Italian banks.


Moreover, preliminary to the multicriteria analysis, we have evaluated the financial soundness of the start-up considered outperforming a scenario analysis on the basis of their Business Plans. The financial analysis has been conducted along two directions. On one hand, a NPV for each scenario has been computed. On the other, financial ratios  have been evaluated and included in the multicriteria analysis.

The paper is organized as follows.
In Section 2, we provide a brief description of ELECTRE-TRI and SMAA-TRI.
In Section 3, we  describe the main phases of the multicriteria rating model proposed to evaluate the riskiness of innovative \textsf{SMEs}. A real case study is carried out in Section 4 to illustrate the judgmental multicriteria credit risk rating presented. Section 5 gives some useful managerial insights on the results obtained by the methodology considered. Finally, some conclusions are included in Section 6.

\section{A sorting model}

In this paper, the multicriteria approach ELECTRE-TRI (\citealp{Yu:1992}) has been proposed as a credit   rating model to assign innovative SMEs into risky categories.
ELECTRE-TRI has been implemented in the framework of SMAA-TRI, to take into account uncertainty and imprecision in its preference parameters. In the next sections, we provide a brief description of ELECTRE-TRI and SMAA-TRI.
\subsection{An overview of ELECTRE-TRI method }

Let $A=\{a_1, a_2, \ldots, a_i, \ldots, a_m \}$ be a  finite set of $m$ alternatives to be assigned 
on the basis of a consistent family of $n$ criteria $G=\{1,2,  \ldots, n\}$  
 to $p$ risk ordered categories $C_p \succ \cdots C_{k}\cdots \succ C_1$, where $C_{k+1}$ consists of a group of  alternatives, better than those in  $C_{k}$.
 
\noindent An alternative $a_i \in A$ evaluated on  the set of  $n$ criteria   is denoted by:

$$a_{i}=(a_{i1}, \cdots, a_{ij}, \cdots a_{in}),$$

\noindent where $a_{ij}$ indicates the evaluation of the alternative $a_i$ on criterion $j$.
 
\noindent Every group of alternatives is separated from the others by means of risk profiles:\\ 
$b_{p-1}, \cdots,b_k, \cdots, b_{1}$. Each profile $b_k$ is the upper limit of the group $k$ and the lower limit of group $k+1$.
For example, considering two  ordered classes, $C_2$ could represent the financed enterprises, while $C_1$  groups the  ones non-funded. As a result,  one risk profile $b_1$  delimits two risk categories to which an enterprise  has to be assigned. The alternatives are assigned to the risk classes on the basis of their comparisons with the risk profiles exploiting the two outranking relations
$a_i S b_k$ and $ b_k S a_i$, which mean, respectively, that alternative $a_i$ is at least as good as  the profile $b_k$ and \emph{vice versa}.

The exploitation of each outranking relation consists of  two phases: the concordance and discordance test. Roughly speaking, the concordance test indicates the level of majority of criteria that supports the outranking relation $a_i S b_k$ (or $b_k S a_i$), while the discordance test represents the strength of the minority of criteria that oppose a veto to the outranking relation $a_i S b_k$ (or $b_k S a_i$).

The concordance test is performed by the computation of the concordance index $C(a_i,b_k)$:

$$C(a_i,b_k)=\sum_{j=1}^{n} w_jc_{j}(a_{ij},b_k),$$

\noindent where $w_j$ indicates the weight of criterion $j$ with the weights summing up to 1 and $c_{j}(a_i,b_k)$ is the partial concordance index with respect to criterion $j$ of the assertion ``$a_i$ is at least as good as the profile $b_k$.'' The partial concordance index $c_{j}(a_{ij},b_k)$ is computed as follows:

$$c_{j}(a_{ij},b_k)=\left\{
\begin{array}{ccl}
  0                                 & \text{if} &  a_{ij}\leq b_{kj}-p_j \\
  \frac{a_{ij}-b_{kj}+p_j}{p_j-q_j} & \text{if} &  b_{kj}-p_j<a_{ij} < b_{kj}-q_j \\
  1                                 & \text{if} &  a_{ij} \geq b_{kj}-q_j,
\end{array}\right.$$

\noindent where $p_j \geq q_j \geq 0$ indicate the preference and indifference thresholds, respectively. Such thresholds take into account the imprecise criteria evaluations.

The preference threshold represents the smallest difference $a_{ij}-b_k$ compatible with the preference of $a_{i}$ on criterion $j$, while the indifference threshold indicates the largest indifference $a_{ij}-b_k$ that preserves indifference between $a_i$ and $b_k$. Similarly, the concordance index $C(b_k,a_i)$ can also be defined.

The second phase consists in the discordance  test, that is performed by the computation of the discordance index $d_{j}(a_{ij},b_k)$  with respect to criterion $j$. Such index is  defined as follows:

$$d_{j}(a_{ij},b_k)=\left\{
\begin{array}{ccl}
  0                                 & \text{if} &  a_{ij}\geq b_{kj}-p_j \\
  \frac{b_{kj}-a_{ij}-p_j}{v_j-p_j} & \text{if} &  b_{kj}-v_j<a_{ij} < b_{kj}-p_j \\
  1                                 & \text{if} &  a_{ij} \leq b_{kj}-v_j,
\end{array}\right.$$

\noindent where $v_j \geq p_j $ is a veto threshold expressing the smallest difference $a_{ij}-b_k$ which is incompatible with the statement $a_i S b_k$.

Finally, a credibility index of the outranking relation is computed by:

$$\sigma(a_{i},b_k)=C(a_{i},b_k)\prod_{j \in \overline{T}}\frac{1-d_{j}(a_{ij},b_k)}{1-C(a_{i},b_k)},$$

\noindent where $\overline{T}$ is the set of criteria such as $d_j(a_{ij},b_k) > C(a_{i},b_k).$
The outranking relation $a_{i}S b_k$  has to be ``defuzzyfied'' on the   the basis of a user-defined threshold $\lambda \in [0.5,1]$.

The relation  $a_{i}S b_k$  is valid if and only if
$\sigma(a_{i},b_k)>\lambda.$ Similarly, the outranking relation $b_k S a_{i} $ is validated if and only if $\sigma(b_k,a_{i})>\lambda.$

Choosing a  value  $\lambda \geq 0.5$,  is equivalent to saying that at least $50\%$ of the  criteria considered are in favour of the assignment of $a_i$ to class $C_k$.

Then, two assignment procedures could be adopted; the pessimistic and the optimistic ones.

The above assignment rules are formulated as follows:

\begin{itemize}
  \item Pessimistic assignment (conjunctive logic): each alternative is compared successively to the profiles $b_{p-1}$, $b_2\ldots b_1$; if $a_i S b_k$, then $a_i$ is assigned to the category $C_{k+1}$, otherwise $a_i$ is assigned to $C_1$, i.e. the worst category;
  \item Optimistic assignment (disjunctive logic): each alternative being compared to the profiles\\
  $b_{1},b_{2}, \ldots, b_{p-1}$; if $b_k S a_i \wedge a_i \neg S b_k$, then  $a_i$ is assigned to the category $C_{k}$ and otherwise $a_i$ is assigned to the group $C_p$, i.e. the best category.
\end{itemize}
Some troublesome situations of incomparability could arise if the two preference statements are verified: $a_i \neg S b_k$ and $b_k \neg S a_i$.

In the present paper, we adopt the pessimistic rule that is  most used in practice,
assigning every alternative $a_i$ to the category $C_{k+1}$ if the following inequalities are verified:

\begin{equation}\label{Sigma}
    \sigma(a_{i}, b_{k}) \geq \lambda \,\,\text{and}\,\,\sigma(b_{k},a_{i}) < \lambda.
\end{equation}

If there aren't veto criteria, it is easy to verify that 
$$\displaystyle \sigma(a_{i}, b_{k})=C(a_{i}, b_{k})=\sum_{j=1}^{n}w_jc_j(a_{i}, b_{k}).$$

As a consequence, the aforementioned  decisional rule (\ref{Sigma}) is simplified as follows:

\begin{equation}\label{simplerule}
    \sum_{j=1}^{n}w_jc_j(a_{i}, b_{k}) \geq \lambda \,\,\text{and}\,\,\sum_{j=1}^{n}w_jc_j(b_{k},a_{i} ) < \lambda.
\end{equation}



\subsection{SMAA-TRI}

In this section, we introduce SMAA-TRI, firstly introduced in \citealp{Tervonen2007} (see also  \citealp{Tervonen2009}).  SMAA-TRI, is a SMAA (Stochastic Multicriteria Acceptability Analysis)  method  that can take into account uncertainty and imprecision on the set of parameters and data required as input by  ELECTRE-TRI.

In this paper,  SMAA-TRI will be used to analyze the robustness of ELECTRE-TRI based on the parameter stability.  By performing Monte Carlo simulations, such  a method generates a set of weights on a $\lambda$ cutting level within an ELECTRE-TRI model. 

The input for SMAA-TRI are the following:
\begin{itemize}
	\item the profiles;
	\item the feasible set of  weights of criteria   defined as:

      $$W=\{w_j \in \rio^+ \colon \sum_{j=1}^{n}w_{j}=1\};$$
      
	\item the $\lambda$ cutting level;
	\item the data and the other parameters of ELECTRE-TRI are supposed deterministic.
\end{itemize}

In \textsf{SMAA-TRI}, a categorization function is defined to evaluate the category $k$ to which an alternative $a_i$ is assigned as follows:

\begin{equation}\label{categoryfnc}
k=F(i, \Delta), 
\end{equation}
where $\Delta$ is the set of parameters of \textsf{ELECTRE-TRI}.

It is also introduced the following category membership function:

\begin{equation}\label{membershipfnc}
m_i^k=\left\{\begin{array}{cl}
1, & \text{if} \,\,F(i,\Delta)=k\\
0, & \text{otherwise}.\\
\end{array}\right.
\end{equation}
  
  The category membership function is used to compute the category acceptability index $\pi_i^k$ that is numerically a
  multidimensional integral over the preference parameter space.
  
  The category acceptability index, generally expressed in percentage-wise,   evaluates the stability of the assignment of an alternative $a_i$  to a category $C_k$. The index is within the range $[0,1]$; if it results 0, this means that the alternative $a_i$ is never assigned to category $C_k$ on the basis of all the parameters randomly generated during the simulations; on the contrary if 1 the alternative $a_i$ is certainly assigned to category $C_k$, considering all the parameters randomly generated during the procedure. In this paper,  \textsf{SMAA-TRI} has been implemented using JSMAA, an open source software in Java (\texttt{www.smaa.fi}; see \citealp{JSMAA}).
  The simulations considered in JSMAA are 10,000.
  
\section{Description of the proposed model }
\label{MCDA}

Presently,  the starting point of the riskiness evaluation process of an  innovative SME is  a Business Plan (BP),
in which the principal used investment appraisal indicators based on cash-flow estimates, like the Net Present Value (NPV) are computed.
In the evaluation process of an innovative SME, such indicators, as it will be shown in the case study, could be useful in a first screening of the enterprise under consideration. 
In the case study, a scenario analysis has been outperformed by considering the base-scenario and two worst scenarios to reject eventual unprofitable projects with NPV$<0$.

Thus, for each enterprise we have evaluated some financial ratios and then we have compared to  the  quartiles values for each sector to which the enterprises, considered in the case study, belong. At the same time, we have also estimated their NPV under each scenario.


Besides the need of a preliminary financial analysis, with respect to  the innovative SMEs, the only evaluation of the financial factors, estimated in the business plan,  is  weakly significant, since if their specific risks  are not analyzed, their economic-financial prospectives could be affected.
 
As a consequence, the problem of sorting  innovative SMEs by intensity of risk  requires to consider several aspects such  as multiple criteria; in particular, the non-financial (qualitative) criteria that are the most relevant ones as it has been pointed out in the introduction.
  
First of all, in a multicriteria problem it is very important to identify all the actors present.  The  principal actor involved in our decision problem is the credit officer of a bank, that is the Decision Maker (DM).
The output of the model will be the sorting of each enterprise to a predefined risk category.

  

Also, in this  decision procedure an important role is also given to the  experts, such as engineers, physicists  or chemists, that   may help the DM    in selecting the proper criteria for evaluating an innovative enterprise, especially for detecting its  technological risks.

In the present paper, in order to evaluate the riskiness of innovative \textsf{SME}s, we adopt the multicriteria model,
\textsf{ELECTRE TRI} (\citealp{Yu:1992}), that belongs to the family of \textsf{ELECTRE} methods, firstly introduced in \citealp{BR68}. Specifically, \textsf{ELECTRE TRI}   assigns an innovative SME to some predefined risk categories by comparing it with some reference profiles delimiting the categories.

\textsf{ELECTRE TRI} is implemented  within the framework of a constructive approach (\citealp{Roy:1993}) since its main operational phases are determined by  an active participation of the DMs where the DMs' preferences are not existent in their minds, but they are revealed during an interactive decision process  to obtain a final recommendation.

%
%
%
%
%
%

The main phases of the  multicriteria  judgmental rating model proposed, i.e. ELECTRE-TRI, are listed hereafter:

\begin{description}

  \item [Phase(i)]  selection of the evaluation criteria;


  \item [Phase(ii)]   definition of the risk classes;

\item [Phase(iii)]   elicitation of importance weights of criteria.

\end{description}

\subsection*{Phase (i): selection of the evaluation criteria}
\label{criteria}

Firstly, in a multicriteria problem  the alternatives have to be evaluated on the basis of  a \emph{consistent} family of criteria (\citealp{Roy:1985}), defined as follows:

\begin{itemize}
  \item two alternatives with the same criteria evaluations have to be assigned to the same risk class (\emph{exhaustive criteria});
  \item an innovative alternative on which a criterion value is decreased in terms of lower risk cannot be assigned to a lower class (\emph{consistent criteria});
  \item an innovative alternative cannot be assigned to a risk class when one criterion is dropped from the family of criteria (\emph{non-redundant criteria}).
\end{itemize}


%
%
%
%
%
%
%

The non-financial criteria, considered in this study, will be gathered in four main groups on the basis of the  risk areas specific of an innovation (see \citealp{Mazzu:2008}).
The risk areas considered are described and listed hereafter.

\begin{itemize}
  \item \textbf{\emph{Development risk.}}
  The enterprise's organization shows a weak  competence to orient itself to  the necessary changes to implement a process or product innovation. Such a risk may be related to some mistakes during the projecting phase of an innovation. Indicators of such a risk may be the management quality, the scientific skills for the patents, or the company profile.
    \item \textbf{\emph{Technological risk.}} Of course, this risk is crucial for any innovative enterprise.
  Its weaknesses in the technological skills  may  result in the failure  of the project. The indicators of such a risk are mainly related to the knowledge of the existing technologies or the pros of the technique considered.

  \item \textbf{\emph{Market risk.}} Such a risk is tied to the continuous  monitoring of the potential market and of the key competitors of the innovation under consideration. For example, its possible indicators may be the adoption of customer evaluation models for the innovation's demand, or knowledge of the competitors.  
  \item \textbf{\emph{Production risk.}} The production of  innovation may be affected by the non-flexible   company structure.
  The production risk indicators may  be the adequacy of the production structure, dependence on the suppliers for raw materials, and the availability of testing and unit pilots.

\end{itemize}

Beyond the qualitative criteria (soft information), we have also taken into account some financial criteria estimated on the basis of the Business Plans presented by the enterprises. 
 
In the paper, we have selected two innovation indicators and three financial ratios.
With respect to the  several innovation indicators available in literature,   we have chosen to adopt the two following innovation indicators: Intangible Assets/Fixed Assets (see \citealp{BankItaly}) and  $R \& D$/Sales (see \citealp{Montfort:2002}). Both indicators are the most used in many empirical studies especially because they are easily measurable.

Concerning the financial ratios, within the multicriteria approach proposed we have considered  ROA (profit before taxes/total assets), Short-term debt/Equity,  and Cash/Total Asset reflecting
the areas of profitability, leverage  and liquidity of each enterprise (see Section 4.1. for a full description of the financial analysis). 

\begin{figure}
\begin{center}

\psset{xunit=0.8cm} \psset{yunit=0.8cm}
\begin{pspicture}[showgrid=false](-5,7)(17,10.5)
\psframe(4,9.5)(7,10.5)

\rput(5.5,10.2){\tiny Set of Criteria}
\rput(5.5,9.8){\tiny $G$}


\psline(5.5,9.5)(5.5,9.2)
\psline(5.5,9.2)(5.5,8.8)
\psline(1,9.2)(1,8.8)
\psline(10,9.2)(10,8.8)
\psline(14.5,9.2)(14.5,8.8)

\psline(-3.5,9.2)(1,9.2)
\psline(1,9.2)(14.5,9.2)

\psframe(-5,8)(-2,8.8)
\psframe(-0.5,8)(2.5,8.8)
\psframe(4,8)(7,8.8)
\psframe(8.5,8)(11.5,8.8)
\psframe(12.5,8)(16.5,8.8)


\psline(-4.5,7.7)(-4.5,7.5)
\psline(-3.7,8)(-3.7,7.7)
\psline(-3,7.7)(-3,7.5)
\psline(-3.5,9.2)(-3.5,8.8)
\psline(1,8)(1,7.5)
\psline(5.5,8)(5.5,7.7)
\psline(4,7.7)(4,7.5)
\psline(7,7.7)(7,7.5)
\psline(9,7.7)(9,7.5)
\psline(10,8)(10,7.7)
\psline(11,7.7)(11,7.5)
\psline(14.5,8)(14.5,7.7)
\psline(12,7.7)(12,7.5)
\psline(13.5,7.7)(13.5,7.5)
\psline(15.5,7.7)(15.5,7.5)
\psline(17,7.7)(17,7.5)

\psline(-4.5,7.7)(-3,7.7)

\psline(4,7.7)(7,7.7)
\psline(9,7.7)(11,7.7)
\psline(12,7.7)(17,7.7)

\rput(-3.5,8.5){\tiny Development Risk }
\rput(-3.5,8.2){\tiny $G^1$ }

\rput(1,8.5){\tiny Technological Risk }
\rput(1,8.2){\tiny $G^2$ }
\rput(5.5,8.5){\tiny Market Risk }
\rput(5.5,8.2){\tiny $G^3$ }
\rput(10,8.5){\tiny Production Risk}
\rput(10,8.2){\tiny $G^4$ }
\rput(14.5,8.5){\tiny Financial criteria}
\rput(14.5,8.2){\tiny $G^5$ }


\psframe(-4.9,7)(-3.9,7.5)
\psframe(-3.5,7)(-2.5,7.5)



\psframe(0.5,7)(1.5,7.5)

\psframe(3.5,7)(4.5,7.5)


\psframe(3.5,7)(4.5,7.5)

\psframe(6.5,7)(7.5,7.5)

\psframe(8.5,7)(9.5,7.5)

\psframe(10.5,7)(11.5,7.5)

\psframe(11.6,7)(12.6,7.5)

\psframe(12.8,7)(13.8,7.5)
\psframe(14,7)(15,7.5)

\psframe(15.2,7)(16.2,7.5)

\psframe(16.4,7)(17.4,7.5)


\rput(-4.4,7.2){\tiny $g^1_{(1)}$}

\rput(-3,7.2){\tiny $g^1_{(2)}$}

\rput(1,7.2){\tiny $g^2_{(3)}$}

\rput(4,7.2){\tiny $g^3_{(4)}$}

\rput(7,7.2){\tiny $g^3_{(5)}$}

\rput(9,7.2){\tiny $g^4_{(6)}$}

\rput(11,7.2){\tiny $g^4_{(7)}$}

\rput(12.2,7.2){\tiny $g^5_{(8)}$}

\rput(13.3,7.2){\tiny $g^5_{(9)}$}

\rput(14.5,7.2){\tiny $g^5_{(10)}$}

\rput(15.7,7.2){\tiny $g^5_{(11)}$}

\rput(16.8,7.2){\tiny $g^5_{(12)}$}

\end{pspicture}
\vspace{0.5truecm}
\caption{A hierarchical family of criteria in the risk evaluation of an innovative SME.}
\label{Hierarchy}
\end{center}
\end{figure}
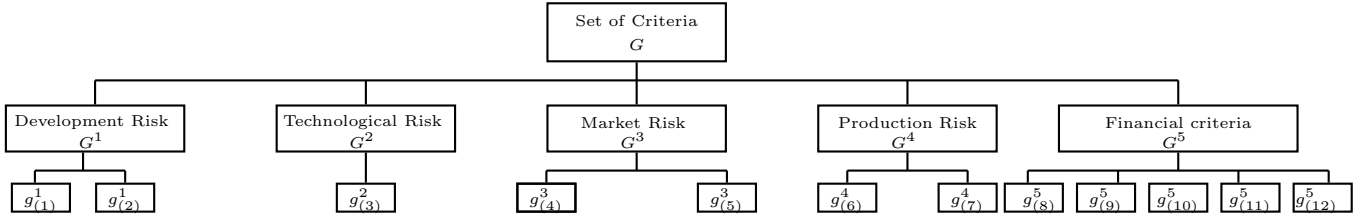

In financing innovative SMEs, we have taken into account  the criteria  hierarchically structured  as follows:

\begin{itemize}

  \item  at the first level, the criteria  have been  denoted by $G^r$  where $r=1,2,3,4,5$  indicate, respectively, the criteria relative to the development ($G^{1}$),  technological  ($G^2$),  market ($G^3$) and  production ($G^4$) risk areas and   financial criteria ($G^5$), i.e. $G=\{G^{1},G^{2},G^{3},G^{4}, G^5\}$ (see figure \ref{Hierarchy}).

  \item at the second level,   the set of  sub-criteria relative to every group of criteria have been considered and  denoted by $G^r=\{g^r_{(1)}, g^r_{(2)},\ldots g^r_{(q)} \ldots, g^r_{(n_r)}\}$ with $q= 1,2, \ldots, n_r$.

 \noindent At this level, we have denoted the set of all  the  sub-criteria by  $G$  where $|G|=n=n_1+n_2+n_3+n_4+n_5$.

\end{itemize}

For the sake of  simplicity, $g^r_{(j)}$ has been used to indicate any  sub-criterion  belonging to $G$  with $j=1,2, \cdots, n$. As is well-known (see \citealp{Figueira:2010}),  the ELECTRE methods require that all the criteria are defined on a common level; henceforth we consider the criteria  at the second level.

\subsection{Phase (ii): definition of the risk classes }
\label{PhaseII}

Within ELECTRE-TRI, the reference profile, delimiting the risk categories, is given by the vector
 $$b_k=\{b_{k1},b_{k2}, \ldots, b_{kj}\},$$

\noindent  $j$ being the criterion index  and whose coordinates indicate the performance of the profile in each of its criteria.  It delimits the lower limit of category $C_{k+1}$ and the upper limit of category $C_k$.
Delimiting the risk classes is also a DM's task.
Determining categories depends on the DM, how (s)he perceives the different criteria evaluated on the basis of the alternatives under consideration.
In the case study presented in Section 4, the loan officers of the Italian bank interviewed have  suggested us to adopt five rating classes expressing a decreasing order of risk from the highest level risk (risk class 1: bad enterprises) to the lowest (risk class 5: very good enterprises).
  
%

%
%


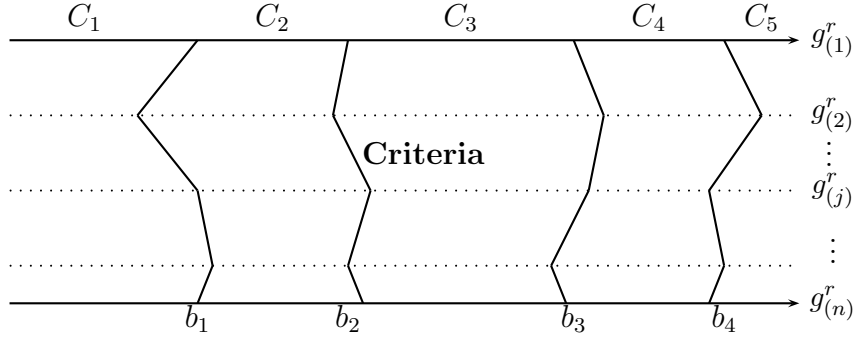
\begin{figure}
\begin{center}
\psset{xunit=1cm} \psset{yunit=1cm}
\begin{pspicture}[showgrid=false](3,3)(14,8)

\psline{->}(2.5,7)(13,7)
\psline{->}(2.5,3.5)(13,3.5)
\rput(3.5,7.3){$C_1$}
\rput(6,7.3){$C_2$}
\rput(8.5,7.3){$C_3$}
\rput(11,7.3){$C_4$}
\rput(12.5,7.3){$C_5$}

\rput(13.45,7){\small $g^r_{(1)}$}

\psline[linestyle=dotted](2.5,6)(13,6)
\rput(13.45,6){\small $g^r_{(2)}$}

\rput(13.4,5.6){$\vdots$}

\psline[linestyle=dotted](2.5,5)(13,5)
\rput(13.45,5){\small $g^r_{(j)}$}

\rput(13.45,4.3){$\vdots$}

\psline[linestyle=dotted](2.5,4)(13,4)
\rput(13.45,3.5){\small $g^r_{(n)}$}

\rput(12,3.3){$b_4$}

\psline(12,7)(12.5,6)
\psline(12.5,6)(11.8,5)
\psline(11.8,5)(12,4)

\psline(12,4)(11.8,3.5)

\rput(10,3.3){$b_3$}

\psline(10,7)(10.4,6)
\psline(10.4,6)(10.2,5)
\psline(10.2,5)(9.7,4)
\psline(9.7,4)(9.9,3.5)

\rput(7,3.3){$b_{2}$}

\psline(7,7)(6.8,6)
\psline(6.8,6)(7.3,5)
\psline(7.3,5)(7,4)
\psline(7,4)(7.2,3.5)

\rput(5,3.3){$b_{1}$}

\psline(5,7)(4.2,6)
\psline(4.2,6)(5,5)
\psline(5,5)(5.2,4)
\psline(5.2,4)(5,3.5)

\rput(8,5.5){\textbf{Criteria}}

\end{pspicture}
\caption{Risk classes defined by the limit profiles.}
\label{categories}
\end{center}
\end{figure}

\subsection{Phase (iii): the importance  weights of criteria}
\label{limit}

One of the most important and difficult tasks of  MCDA is  the determination of importance weights of criteria. In  literature, several methods have been proposed to assess  weights of criteria (see, for example \citealp{Bana:1994}, \citealp{Bana:1997}, \citealp{Lagreze} and \citealp{Mousseau:2001}).

In addition to  the different  preference models of elicitation of weights,  an important question in MCDA is the relative importance of criteria (see \citealp{Mousseau:1995} and \citealp{Roy:1996}).
Indeed, there is a different meaning between weights in compensatory methods, like the Multi-Attribute Utility Theory (MAUT)  (\citealp{Keeney:1976}) and non-compensatory methods like ELECTRE (\citealp{Roy:1991}) and PROMETHEE  (\citealp{Brans:1984}).
In compensatory methods, weights of criteria are essentially trade-offs between criteria and are constants dependent on their scale and range.
On the other hand,  in non-compensatory methods the weights of criteria have  an intrinsic value and do not change on the basis of  the scale of criteria used.

Among these, we recall    the revised Simos' method (\citealp{BRJF02}), known also as the \emph{cards method}, that has been used in the case study illustrated in the next section.

Within the Simos' procedure, the DM ranks the criteria from the least to the most important by using a set of cards (one for each criterion). The DM can also insert some blank cards to separate the relative importance between criteria.
If the DM inserts no white cards between criteria, this means that these criteria have not the same weights and their relative difference can be taken as unit measure $u$  for weights.
With a similar reasoning, if the DM inserts one white card, two white cards, etc, this means respectively a difference of weights of two times $u$, three times $u$, etc.

In the next section, where a case study is presented, the \emph{cards method} has been adopted to obtain ordinal information on the weights of the criteria and then to obtain indirectly the weights of criteria from the DM's ranking.

Within the case study, we have considered different  loan officers as it is supported by the literature that has investigated the role of the loan officers in  lending decision making in banks (see for example \citealp{Hemlin:2014} 
that examines the organizational factors that influence bank loan officers).
Specifically, we have interviewed five loan officers of one of the main Italian banks, which have given different orders of importance of weights of criteria depending on their  preference attitudes.

In  Section \ref{simos}, a case study has been carried by taking into account the different DMs interviewed.  For each DM, we have applied \textsf{SMAA-TRI} by varying the $\lambda$-cutting level represented by a  stochastic variable uniformly distributed. Then, to take into account simultaneously  all the DMs' sets of weights, the evaluation of each criterion ranges in an interval delimited by the maximum and the minimum weight for all DMs. Since the sets of  weights are imprecise, we have outperformed again \textsf{SMAA-TRI}, computing the category acceptability indices for each alternative.


\section{Start-up case study}

In this section, we carry out a real case study regarding four innovative SMEs, headquartered in Italy, that we have contacted asking them  their business plans and to  fill out a  structured questionnaire. The data collected in  the questionnaire concern some  information on the company, distribution and customer networks, demand
forecasting, supply chain information, the owners' Company CV and eventual  awards  received by the company.
For the sake of privacy, we have renamed the companies as A, B, C and D. Moreover, we have interviewed five loan officers of one of the main Italian banks to simulate  the financing of such companies on the basis of the multicriteria rating model proposed in Section \ref{MCDA}.
Hereafter, we report a brief description of the considered companies.
\vspace{0.1cm}

\noindent\textbf{Companies}

\vspace{0.1cm}
\begin{enumerate}

\item Company A is  a biotechnological  start-up, operating in the field of the green economy. It  has developed some biological systems based on plants and micro-organisms forming some eco-friendly barriers,  to prevent soil from the hydrological destruction and the environmental pollution. The services provided by it, consist of realizing such biological barriers, also  giving consultancy; furthermore its services are offered at  lower prices than the ones of the traditional techniques.

\item Company B is a technological company   with a strong expertise in digital communications.
The  innovative idea of the company is to develop a ``Water-MeMo'' that is a wireless sensor network for water leakage detection, based on energy harvesting. In particular, its technology can be applied not only in the water distribution field, but also in other energy applications, such as heat and gas, or environmental monitoring.
Its main innovation is   in  the green technology FluE (Fluid Energy) that is based on some  sensors  auto-charged by the  energy produced by the fluid itself trough some piezoelectric foils.

\item Company C is a high-technological start-up, an R $\&$ D  mechanical design and service provider company with a focus on material recovery systems applied to thin-film deposition processes. Its mission is providing breakthrough technology, in order to increase the efficiency of PVD (Physical Vapour Deposition) processes, today used to produce microchips, MEMS (Micro Electro-Mechanical Systems), solar cells and other hi-tech devices.


\item Company D is a high-technological start-up that aims to produce and commercialize graphene and carbon nanotubes for industrial use and for research, to develop new materials based on nano-particles and provide technical assistance and advice to businesses willing to use these technologies. The company is mainly focused in producing nano-engineered epoxy resins used in the manufacture of sport equipment, for example sailing boats or kiteboards.

\end{enumerate}

\subsection{Financial analysis  }

\label{FinancialAnalysis}
The starting point of the financial analysis, presented in this section,  is  the  enterprises' business plan, on which we have out-performed  a scenario analysis.  From   the base-case scenario,  we have developed two different worst scenarios considering the cash-flows lowered by 20$\%$ and 40$\%$.

First of all, we have evaluated  the NPVs for each scenario since this is a screening tool to know if the projects are profitable.
The NPV under the different scenarios have been computed by using a risk free  rate of $7.93\%$, commensurated with the performance of Italian public debt securities in date 30/9/2012 (Base informativa pubblica, Banca d'Italia, 2013).
The results are reported for each enterprise in  Table \ref{Unica} of  the Appendix A. 
In all the enterprises under each scenario, all the NPVs result positive and consequently all the projects are profitable. Then,  we have computed some financial ratios. 

First of all, let us remark that the  enterprises have considered different planning horizons; Company D has presented a three years BP, Company A  a four year BP, Company B has considered a five years BP and Company C a six years BP.
Even if track records of innovative firms are not available, we have evaluated some financial ratios of the enterprises of the case study  on the basis of their business plan.
On the basis of the available financial data,  the following financial ratios have been selected: ROA, Short-term debt/Equity,  and Cash/Total Asset, reflecting respectively
the areas of profitability, leverage  and liquidity of each enterprise (these ratios have been already selected to evaluate the US SMEs in the paper of \citealp{AltmanSabato2005}).

To the aim of detecting   the  Italian firms in the same sector of the four enterprises under evaluation, we have considered their ATECO codes which are used to classify companies based on their activities. 

From AIDA dataset,  we have reported the financial ratios of the balance-sheets  of the  sectors of  companies A, B, C and D respectively,  composed of 515, 13,735, 1,001, and 1,122 SMEs  (enterprises with a number of employees  lower than 250) over the period 2008-2012.
  
Finally,  from 2008 to 2012 
we have evaluated  the quartiles of each financial ratio for all the enterprises in the sample. The quartiles  with respect to each sector has been assumed as the  limit profiles in performing the multicriteria model of ELECTRE-TRI (see Table \ref{financialindicators}).




\subsection{Phase (i): building a family of criteria}

Since most of the criteria, considered in this paper,  are qualitative, 
for a few of them it has been natural to adopt   a  binary codification while for most of them, 
for  simplicity,   we have used the same  ordinal scale on a five point scale.
Let also  note that all the criteria evaluations are increasing, i.e. the more the better excepting the Short-term debt/Equity that has to be minimized.

 
The  family of criteria have been elaborated by the credit officers interviewed with the help of the experts. Even more on the basis of the BP presented by the firms, the criteria evaluations have been given by the credit officers in cooperation with the experts that can detect the specific risks of each enterprise. The criteria considered in the case study are described hereafter (see also Table \ref{Hof} for an overview of the criteria).\\


\textbf{Criteria related to the development risk.	}

\begin{description}

%

%
%

\item[[$g^1_{(1)}\!\!\!$]] Awards:  a five-points scale is adopted, ranging from one point, if a company hasn't received no award, to five points, if the innovative enterprise has received an international award. 


\noindent The idea of Company A has been certified by different awards.
Among others the  D2T Start Cup given by  Trentino Sviluppo and it  has also been classified first in the competition AGRIstart Up. On the basis of such criterion,  the DMs suggest an evaluation of  four points.

\noindent Company B has obtained the following awards: research grant Working Capital 2011, seed fund eCapital 2011 and ItaliaCamp 2012. On the basis of such criterion,  four points have been given to Company B.

\noindent Even Company C has received different awards: both Italian and international. 
Among others Company C has obtained the  third place in the Innovact Campus Award 2012 in Reims (France).
On the basis of such criterion,  five points have been given to Company C.

\noindent Company D  won  the  Business  Plan  Competition  eCapital  2010,
competition  aimed  at  creating  innovative  startups.  In  2011, it  was
finalist  at  the Italian National  Innovation  Award  ``PNI  -  Telecom
Working Capital''. On the basis of such criterion,  four points have been given to Company C.

\item [[$g^1_{(2)}$\!\!\!]] Scientific skills: a five-points scale is adopted, ranging from one point, if the principal partners have no specific studies, to five points, if the owners of the innovative enterprise have received a PHD and have some specific job experiences. 

\noindent Company A. The three principal partners are agronomists and are specialized in microbiology and in plant genetics. Company A has received three points on this criterion.

\noindent Company B is composed of three young engineers, a full professor, a strategic marketing expert and ArieLab s.r.l., a spin-off of the University Politecnica delle Marche. Company B has received five points on this criterion.

\noindent Company C. The management team is composed of a well qualified staff. For example, the CEO has a Master degree  in Engineering, or the CTO has a Master degree in Physics and NanoScience. The DMs have given three points
on this aspect.

\noindent Company D. All the five members are well qualified; some of them have  a degree in engineering, others are specialized in nanotechnology and one of them is specialized in sailing prototypes that are the main products in which the company is investing its research and sale activities. The DMs have given five points
on this criterion.

%
%
%

\end{description}

\textbf{Criteria related to the technological risk.}

\begin{description}


\item[[$g^2_{(3)}\!\!\!$]] Pros of the techniques used in comparison to other similar already existing: a five-points scale is adopted;  one point means that the advantages are irrelevant,  while  five point levels are given if they are significant.

\noindent Company A has received three points, since there are  other methods that can be used to prevent soil from the hydrological destruction and the environmental pollution.

\noindent On the basis of this criterion, Company B has received   five points  since the only similar existing technique consists of putting a few Data Loggers  in some points of the water pipelines to register the noise by a microphone.
Anyway, such products are too expensive and detect the soundness of the water distribution \textit{una tantum}.

\noindent On the basis of this criterion, Company C has received   five points  since other existing techniques need too complex chemical processes to recover precious materials.

\noindent Company D  has received  three point levels, 
since the retail prices of their epoxy resins are too high in comparison to  other existing ones.

\end{description}

\textbf{Criteria related to the market risk.}

\begin{description}
\item[[$g^3_{(4)}\!\!\!$]] Presence of key competitors: a five-points scale is adopted, ranging from one point, if there is  a monopolist competitor in the market, to five points, if the innovative enterprise is the only one competitor present in the market.

\noindent Company A has one competitor, that adopts a methodology much more expensive and with many different drawbacks, for example the plants used are not very suitable for the soil. Consequently, on the basis of such criterion Company A has obtained  an evaluation of  four points.

\noindent Company B doesn't have competitors in the market. The only similar solution, i.e. the Data Logger has many drawbacks. The DMs suggest an evaluation of five points on this criterion.

\noindent   Even if Company C has registered a patent, this doesn't represent a barrier to avoid that other enterprises could enter the market, by slightly modifying the system developed by Company C.
Some possible similar solutions could be: vacuum ad hoc chambers, static monitors and mechanical removing.
Anyway, the above techniques need the support of complex chemical processes. On the basis of such criterion, 
The DMs suggest an evaluation of  two points.

\noindent Company D  has different competitors in nanotechnology even if operating with different prices and quality in the their products.  The DMs give an evaluation of two points  on the basis of this criterion.

\item[[$g^3_{(5)}\!\!\!$]] Potential market: a five-points scale is adopted; a one point means that the market is reducing while 5 means that the market is booming.

\noindent  Company A's potential market is limited  by the traditional techniques, mainly based on chemical products, preferred to the innovative solution offered by Company A.  In fact, there is a cultural resistance to accept this new type of methodology to prevent  soil from the hydrological destruction and the environmental pollution, even if nowadays there is much more attention to environmental problems. On the basis of this criterion, a prudential evaluation of three has been considered.

 \noindent With respect to this point, the DMs believe that the Company B's market is reducing (1 point). More precisely, even if the private and public companies running the water distribution in Italy need a system to monitor the water leakage, they are not financially sound. 

\noindent Company C. The  market of thin film deposition is booming, with an estimation of market value amounting to 22.4 billion USD in 2015. The use of precious materials (Au, Ag, Pt and Pd) is very common in optics and electronics. On this criterion, Company C has received an evaluation of five points.

\noindent Company D.   Within  the nanotechnology market  nanomaterials  are  a  booming business  with  benefits  for  all  other  sectors.  Carbon  nanotubes
are  in  this  case,  together  with  some  materials  connected  to  them
(e.g. graphene and fullerenes),  the  most  important  products  of  the entire  sector. On this criterion, Company D  has received an evaluation of five points.

%
%
%
%
%
%
%
%
%
%

\end{description}

\textbf{Criteria related to the production risk.}

\begin{description}
\item[[$g^4_{(6)}\!\!$]] Availability of testing and unit pilots: codes yes (1) or no (0).

\noindent  Company A has not yet realized any pilot installations that can be useful to show to the potential clients the advantages of the technique proposed compared to the traditional ones.
 
\noindent Company B hasn't yet produced any unit pilots, even if it is one of its immediate projects.

\noindent Company C has developed a first prototype and tested it in the vacuum chamber of a thermal evaporator at the laboratory of CNR.

\noindent The Company D's has started   a  pilot
plant  for  the supplement of  nanocarbons  filler  to  thermosetting
epoxy  resins.

\item[[$g^4_{(7)}\!\!\!$]] Owner of a patent: codes yes (1) or no (0).

\noindent Company A hasn't yet applied for a patent even if it is one of its projects.

\noindent Company B hasn't yet registered the patent for the FluE sensor, even if it is one of its goals.

\noindent The Company C  has a right of an Italian patent entitled: \textit{Palette system for the recovery of metals from thin-film deposition facilities} and also of an International application entitled: \textit{Palette modular device for collection of metals in this film deposition equipment}.

\noindent The Company D's  hasn't yet applied for a patent, but it aims to
record  strategic  patents  on  its  core  product  and  a  series  of
multiple  patents  that  go  to  cover  similar  areas  of  research  and
development  thus  protecting  future  market  segments.
\end{description}

\textbf{Financial criteria}

\begin{description}

\item[[$g^5_{(8)}\!\!$]] Intangible Assets/Fixed Assets.

\item[[$g^5_{(9)}\!\!$]] $R \& D$/Sales.

\item[[$g^5_{(10)}\!\!$]] ROA.

\item[[$g^5_{(11)}\!\!$]] Short-term debt/Equity.

\item[[$g^5_{(12)}\!\!$]] Cash/Total Asset.

\end{description}

\begin{table}[!h]
\centering
\caption{Family of criteria\label{Hof}}
\resizebox{8cm}{!}{\begin{tabular}{|l|l|}
\hline
              $\mathbf G^1$: \small{\textbf{development risk}   }          &     \small{\textbf{Codes}}   \\
  \hline
  \hline
                                                         
   \hline
                                                          
     $g^1_{(1)}$:  \small{awards}  &  no awards (1) \\
		                               &  municipal (2) \\
																	 &  regional (3) \\
																	 &  national (4)\\
																	 &  international (5)  \\
     \hline
     $g^1_{(2)}$:   \small{scientific skills} &  no skills (1) \\
		                               &  degree (2) \\
																	 &  master (3) \\
																	 &  PHD (4)\\
																	 &  PHD+Work experiences (5)  \\
\hline
    $\mathbf G^2$: \small{\textbf{technological risk}} &  \small{\textbf{Codes}}  \\
  \hline
  
          \hline
                                                             &      irrelevant  (1)      \\
                                             &      weakly significant (2)      \\
     $g^2_{(3)}$:            \small{ pros of the technique}    &  significant (3)  \\
                                                      & strongly significant (4)\\
                                                                & very significant (5)\\
         
          \hline
        $\mathbf G^3$: \small{\textbf{market risk }} &  \small{\textbf{Codes}} \\
  \hline
     
                                                         &      monopolist  (1)      \\
                                             &      numerous competitors  (2)      \\
     $g^3_{(4)}$:            \small{ key competitors}    &  few competitors (3)  \\
                                                      & one competitor (4)\\
                                                                & start-up (5)\\
     
   \hline
                                                                      &  reducing (1)  \\
                                                                        &   static (2) \\
      $g^3_{(5)}$:       \small{potential market }                       & weakly rising (3)\\
      
                                                          & rising (4) \\
                                                          & booming (5)      \\
     
\hline
    $\mathbf G^4$: \small{\textbf{production risk}} &  \small{\textbf{Codes}} \\
  \hline
  
  $g^4_{(6)}$:  \small{availability of testing}    &  no (0),  yes (1)   \\
             \small{and unit pilots}  & \\
             \hline
  $g^4_{(7)}$:  \small{owner of a patent}    &  no (0),  yes (1)   \\
             
            \hline
   $\mathbf G^5$: \small{\textbf{Financial criteria}} &  unit \\
  \hline
  $g^5_{(8)}$:    intangible assets/fixed assets &  percentage   \\
  \hline
  $g^5_{(9)}$:            $R \& D$/sales    &  percentage  \\
              \hline 
              $g^5_{(10)}$: ROA  & percentage\\
              \hline
              $g^5_{(11)}$: Short term debt/Equity & percentage \\
              \hline
             $g^5_{(12)}$: Cash/Total Asset & percentage\\   
              \hline
       \end{tabular}}
\end{table}

Summing up, for each enterprise the  scores on the considered criteria are showed in Table \ref{EvMatrix}.

\begin{table}[htbp]
  \centering
  \caption{Evaluation matrix.}
  \resizebox{16cm}{!}{  \begin{tabular}{|c||rrrrrrrrrrrr|}
   \hline
  Company    & $g^1_{(1)}$  & $g^1_{(2)}$ & $g^2_{(3)}$ & $g^3_{(4)}$ & $g^3_{(5)}$  & $g^4_{(6)}$ & $g^4_{(7)}$ & $g^5_{(8)}$ & $g^5_{(9)}$ & $g^5_{(10)}$ &  $g^5_{(11)}$ & $g^5_{(12)}$\\  
    \hline
    \hline
  A&  4     & 3     & 3     & 4     & 3     & 0     & 0     & 0.55  & 0.06  & 0.24  & 0.18  & 0.74 \\
  B&  4     & 5     & 5     & 5     & 1     & 0     & 1     & 0.72  & 0.17  & 0.03  & 0.12  & 0.51 \\
  C &  5     & 3     & 5     & 2     & 5     & 1     & 1     & 0.18  & 0.05  & 0.94  & 0.3   & 0.56 \\
  D &  4     & 5     & 3     & 2     & 5     & 1     & 0     & 0.06  & 0.14  & 0.52  & 0.11  & 0.26 \\
   \hline
   \end{tabular}}%
  \label{EvMatrix}%
\end{table}%

\subsection{Phase (ii): determining the limit profiles}

The limit profiles are built by the DMs who determine  the performance on each criterion with respect to every category.
The DMs have considered four  common limit profiles with respect to the qualitative criteria (see Table \ref{profiles}).

\begin{table}[htbp]
  \centering
  \caption{Limit profiles with respect to the qualitative criteria.}
  \begin{tabular}{|r||rrrrrrr|}
    \hline
     & $g^1_{(1)}$  & $g^1_{(2)}$ & $g^2_{(3)}$ & $g^3_{(4)}$ & $g^3_{(5)}$ & $g^4_{(6)}$ & $g^4_{(7)}$  \\
    \hline
    \hline
 $b_1$  &     1     & 1     & 1     & 1     & 1     & 0     & 0 \\
 $b_2$  &     2     & 2     & 2     & 2     & 2     & 0     & 0 \\
 $b_3$  &     3     & 3     & 3     & 3     & 3     & 0     & 1 \\
 $b_4$  &     4     & 4     & 4     & 4     & 4     & 1     & 1 \\
   \hline
    \end{tabular}%
  \label{profiles}%
\end{table}%

Instead, concerning the financial criteria, they have selected a different set of limit profiles for each enterprise under consideration (see Table \ref{ProfilesFinancial}).
Such choice has been justified by considering that each enterprise belongs to a  sector different from the others. Every  sector is characterized by a specific financial risk.
As explained in Section \ref{FinancialAnalysis}, we have considered the   Italian firms in the same sector of the four enterprises under evaluation on the basis of their ATECO codes.  
 We have calculated the quartiles of each financial criterion for all the enterprises in the sample. The quartiles  with respect to each sector have been adopted as the  limit profiles in performing the multicriteria model of ELECTRE-TRI. 
 The upper bound of the less risky category $C_5$  has been estimated by considering the maximum of every criterion with respect to all sectors.
 Only the limit profiles for R$\&$D have been elaborated by the DMs and are the same for all the sectors.

\begin{table}[htbp]
  \begin{center}
\caption{Limit profiles with respect to the financial criteria\label{ProfilesFinancial}}
\subtable[Company A\label{ProfilesA}]{
    \begin{tabular}{|r||rrrrr|}
    \hline
            &$g^5_{(8)}$ & $g^5_{(9)}$ & $g^5_{(10)}$ &  $g^5_{(11)}$ & $g^5_{(12)}$\\ 
            \hline
            \hline
    $b_1$   &0     & 0.03  & $-0.03$ & 5.44  & 0.02 \\
    $b_2$   &0.01  & 0.05  & 0.01  & 1.42  & 0.07 \\
    $b_3$   &0.2   & 0.07  & 0.05  & 0.14  & 0.18 \\
    $b_4$   &1.34  & 0.1   & 0.1   & 0.14  & 0.21 \\
    \hline
    \end{tabular}}\quad
\subtable[Company B\label{ProfilesB}]{
    \begin{tabular}{|r||rrrrr|}
    \hline
          &$g^5_{(8)}$ & $g^5_{(9)}$ & $g^5_{(10)}$ &  $g^5_{(11)}$ & $g^5_{(12)}$\\ 
            \hline
            \hline
   $b_1$   & 0     & 0.03  & $-0.01$ & 3.72  & 0.01 \\
   $b_2$   & 0.17  & 0.05  & 0.03  & 1.22  & 0.06 \\
   $b_3$   & 1.34  & 0.07  & 0.09  & 0.31  & 0.16 \\
   $b_4$   & 1.34  & 0.1   & 0.1   & 0.14  & 0.21 \\
    \hline
    \end{tabular}} \quad
     \subtable[Company C\label{ProfilesC}]{
    \begin{tabular}{|r||rrrrr|}
    \hline
           &$g^5_{(8)}$ & $g^5_{(9)}$ & $g^5_{(10)}$ &  $g^5_{(11)}$ & $g^5_{(12)}$\\ 
            \hline
            \hline
  $b_1$   &  0     & 0.03  & $-0.01$ & 3.14  & 0.01 \\
  $b_2$   &   0.09  & 0.05  & 0.04  & 1.07  & 0.06 \\
  $b_3$   &  0.43  & 0.07  & 0.1   & 0.28  & 0.16 \\
  $b_4$   &  1.34  & 0.1   & 0.1   & 0.14  & 0.21 \\
     \hline
    \end{tabular}} \quad
    \subtable[Company D\label{ProfilesD}]{
    \begin{tabular}{|r||rrrrr|}
    \hline
           &$g^5_{(8)}$ & $g^5_{(9)}$ & $g^5_{(10)}$ &  $g^5_{(11)}$ & $g^5_{(12)}$\\ 
            \hline
            \hline
  $b_1$   &  0     & 0.03  & $-0.04$ & 2.55  & 0.03 \\
  $b_2$   &  0.07  & 0.05  & 0     & 0.67  & 0.08 \\
  $b_3$   &  0.91  & 0.07  & 0.04  & 0.14  & 0.21 \\
   $b_4$   &  1.34  & 0.1   & 0.1   & 0.14  & 0.21 \\
 \hline
    \end{tabular}}
       \end{center} 
\end{table}%

For simplicity, the preference and indifference thresholds have been set equal to zero.

\subsection{Phase (iii): eliciting the set of weights}

\label{simos}

Let us assume a  total preorder $\preceq$, i.e., a reflexive binary relation
on the set of criteria $G$ satisfying two additional properties: transitivity and  completeness.
In this case, the preference relation can be decomposed into its symmetric part  $\sim$, called \textit{indifference}, and into its asymmetric part $\prec$, called \textit{strict preference}, whose semantics are, in a multicriteria context, respectively:

\begin{itemize}

	\item $g_{j_{1}}^{(r_1)}\sim g_{j_{2}}^{(r_2)}\,\,\Leftrightarrow\,\,g_{j_{1}}^{(r_1)}\,\,\text{is equally important to}\,\, g_{j_{2}}^{(r_2)},$

	\item $g_{j_{1}}^{(r_1)}\prec g_{j_{2}}^{(r_2)}\,\,\Leftrightarrow\,\,g_{j_{1}}^{(r_1)}\,\,\text{is less important than}\,\, g_{j_{2}}^{(r_2)},$
\end{itemize}

where $g_{j_{1}}^{(r_1)}$ and $ g_{j_{2}}^{(r_2)}$ are two distinct criteria from the set $G$.


In the following, we report the preference orders of five   credit officers  of one of the main Italian bank interviewed within the case study.

For the first DM (DM1, say), the most important criterion assumed in the evaluation of an innovative enterprise has been the existence of a unit pilot and the second most important criterion has been the patent.

DM1 has given this preference information on the criteria:

 $$g^{1}_{(1)} \sim g^{1}_{(2)} \prec g^{5}_{(8)} \sim g^{5}_{(9)}\prec  g^{3}_{(4)} \sim g^{3}_{(5)}   \prec g^{5}_{(10)} \sim g^{5}_{(11)} \sim g^{5}_{(12)}\prec g^{2}_{(3)}\prec g^{4}_{(7)}    \prec g^{4}_{(6)} .$$

The second DM (DM2, say) has expressed  these preference statements:

$$g^{4}_{(7)} \prec g^{5}_{(8)} \sim g^{5}_{(9)} \prec g^{1}_{(1)}\sim g^{1}_{(2)} \prec g^{5}_{(10)} \sim g^{5}_{(11)}\sim g^{5}_{(12)} \prec g^{2}_{(3)} \prec g^{4}_{(6)} \prec g^{3}_{(4)} \prec g^{3}_{(5)}.$$

In this case, the  prevailing criteria in evaluating an innovative enterprise have been the potential market and the presence of key competitors, i.e. the sub-criteria relative to the risk market.

The DM3's preference information has been the following:

$$ g^{3}_{(4)} \sim g^{3}_{(5)}  \prec g^{5}_{(8)} \sim g^{5}_{(9)} \prec g^{4}_{(7)} \prec g^{4}_{(6)} \prec g^{1}_{(1)}\sim g^{1}_{(2)} \prec g^{5}_{(10)} \sim g^{5}_{(11)} \sim g^{5}_{(12)} \prec g^{2}_{(3)}.$$

In this case, the most important weights considered by the DM3 have been the sub-criteria relative to the technological risk and the second ones the financial ratios.

The DM4's preference information has been given by:

$$  g^{5}_{(8)} \sim g^{5}_{(9)}\prec g^{1}_{(1)}\sim g^{1}_{(2)}\prec g^{3}_{(4)}  \prec g^{4}_{(6)} \prec g^{5}_{(10)} \sim g^{5}_{(11)} \sim g^{5}_{(12)} \prec g^{2}_{(3)}\prec g^{3}_{(5)}   \prec g^{4}_{(7)}.$$
In this case, the most important criterion is assumed to be the patent's ownership and the second  most important one is the potential market.
The DM5's preference information has been:

$$  g^{1}_{(1)} \sim g^{1}_{(2)}\prec g^{4}_{(6)}
\prec g^{3}_{(4)} \prec g^{3}_{(5)} \prec g^{4}_{(7)} \prec g^{2}_{(3)} \prec g^{5}_{(8)} \sim g^{5}_{(9)} \prec g^{5}_{(10)}\sim g^{5}_{(11)}   \sim g^{5}_{(12)}.$$
In the last case, the most important criteria have been the financial ones.

In all the five DMs, the criteria    $(g^{1}_{(1)}, g^{1}_{(2)})$,   $(g^{5}_{(10)},g^{5}_{(11)})$ and $(g^{5}_{(13)},g^{5}_{(14)},g^{5}_{(15)})$ have been considered equally important.

As explained in Section 3.2, the method adopted to assess the criteria weights is the Simos' procedure (the computations relative to the DM1's  weights  are reported in the Appendix).
For all the DMs, the criteria weights obtained are showed in Table \ref{weights}.

\begin{table}[htbp]
  \centering
  \caption{Weights for each DM.}
  \resizebox{16cm}{!}{  \begin{tabular}{|c||cccccccccccc|}
    \hline
  & $g^1_{(1)}$  & $g^1_{(2)}$ & $g^2_{(3)}$ & $g^3_{(4)}$ & $g^3_{(5)}$  & $g^4_{(6)}$ & $g^4_{(7)}$ & $g^5_{(8)}$ & $g^5_{(9)}$ & $g^5_{(10)}$ &  $g^5_{(11)}$ & $g^5_{(12)}$\\  
    \hline
  DM1&     0.0250 & 0.0250 & 0.1650 & 0.0560 & 0.0560 & 0.1960 & 0.1810 & 0.0400 & 0.0400 & 0.0720 & 0.0720 & 0.0720 \\
DM2&    0.0530 & 0.0530 & 0.0934 & 0.1740 & 0.1840 & 0.1640 & 0.0230 & 0.0330 & 0.0330 & 0.0632 & 0.0632 & 0.0632 \\
DM3&     0.1120 & 0.1120 & 0.1390 & 0.0190 & 0.0190 & 0.0720 & 0.0600 & 0.0460 & 0.0460 & 0.1250 & 0.1250 & 0.1250 \\
DM4&     0.0330 & 0.0330 & 0.1490 & 0.0540 & 0.1600 & 0.0640 & 0.1700 & 0.0230 & 0.0230 & 0.0970 & 0.0970 & 0.0970 \\
DM5&    0.022 & 0.022 & 0.1   & 0.061 & 0.074 & 0.035 & 0.087 & 0.112 & 0.112 & 0.125 & 0.125 & 0.125 \\
\hline
    \end{tabular}}%
  \label{weights}%
\end{table}%

\subsection{Computing the category acceptability indices}
 
In this section, we have computed 
the category acceptability indices, according to the different DMs, by using the JSMAA software (the results are showed in  Table \ref{Category}). In all the computations, the $\lambda$-cutting level has been represented by  a stochastic variable uniformly distributed in the range $[0.65,0.85]$.

\begin{table}[htbp]
\begin{center}
\caption{Category acceptability indices according to the different DMs.\label{Category}}
\subtable[DM1\label{CatAccDM1}]{\begin{tabular}{|c||ccccc|}
\hline
	Company	  & $C_1$ & $C_2$ & $C_3$ & $C_4$ & $C_5$\\
		  \hline
 A &   0\%     & 0\%     & 36\%     & 64\%     & 0\% \\
  B  &     0\%    &  0\%     &  0\%    &  74\%     & 26\% \\
  C  &   0\% & 0\% & 0\% & 11\%     & 89\% \\
  D  &   0\%     & 0\%     & 63\%     & 37\%     & 0\% \\
			\hline
		\end{tabular}}
		\quad \quad \subtable[DM2\label{CatAccDM2}]	{\begin{tabular}{|c||ccccc|}
		\hline
		  Company & $C_1$ & $C_2$ & $C_3$ & $C_4$ & $C_5$\\
		  \hline
     A & 0\%     & 0\%    & 0\%  & 100\%  & 0\% \\
      B  &   0\%     &  17\%    &18\%     &  65\%    & 0\%\\
		 C   & 0\% & 0\% & 46\% & 31\%     & 23\% \\
      D  &   0\%     & 40\%     & 47\%     & 13\%     & 0\% \\
    	\hline
		\end{tabular}}
		\quad \quad \subtable[DM3\label{CatAccDM3}]{\begin{tabular}{|c||ccccc|}
		\hline
		 Company & $C_1$ & $C_2$ & $C_3$ & $C_4$ & $C_5$\\
		  \hline
        A & 0\%     & 0\%     & 0\%  & 100\%  & 0\% \\
        B  &  0\%      & 0\%     &  0\%   & 0\%     & 100\%\\
			  C   & 0\% & 0\% & 0\% & 46\%     & 54\% \\
        D &   0\%     & 0\%     & 0\%     & 57\%     &43\% \\
   \hline	\end{tabular}}
		\quad \quad \subtable[DM4\label{CatAccDM4}]{\begin{tabular}{|c||ccccc|}
		\hline
		 Company & $C_1$ & $C_2$ & $C_3$ & $C_4$ & $C_5$\\
		  \hline
      A & 0\%     & 0\%     & 23\%  & 77\%  & 0\% \\
      B  &    0\%    &  5\%   & 14\%    &  33\%    & 48\% \\
      C   & 0\% & 0\% & 0\% & 0\%     & 100\% \\
      D  &   0\%     & 0\%     & 49\%     & 51\%     & 0\% \\
			\hline
		\end{tabular}}\quad \quad \subtable[DM5\label{CatAccDM5}]{\begin{tabular}{|c||ccccc|}
		\hline
		 Company & $C_1$ & $C_2$ & $C_3$ & $C_4$ & $C_5$\\
		  \hline
      A & 0\%     & 0\%     & 27\%  & 73\% &0\% \\
      B  &   0\%     &  0\%    &  22\%   &  18\%    & 60\% \\
      C   & 0\% & 0\% & 69\% & 20\%     & 11\% \\
      D  &   0\%     & 0\%     & 55\%     & 45\%     & 0\% \\
			\hline
		\end{tabular}}
		\end{center}
				\end{table}
\subsection{Category results}

In Table \ref{Results-Sum-up}, we have reported  the highest category acceptability  for each DM obtained during the simulations. One can notice that the assignment of Company A to the category $C_4$ is very stable since all the DMs agree with the same assignment.  
Company B is   assigned  to category $C_5$ by all the DMs except for the DM1. So alternative $B$ has been sorted in category $C_5$.
Company C is assigned to category $C_5$ by three DMs while the highest category acceptabilities for DM2 and DM5 are obtained, respectively, for the classes $C_4$ and $C_3$ even if DM2 and DM5 have a non zero probability of assignment of Company C to $C_5$ equal, respectively, to 23$\%$ and 11$\%$. Thus, alternative C has been assigned to category $C_5$.

Company D is is assigned to category $C_3$ by three DMs, since DM3 and DM4 slightly disagree on this assignment.
But the DM4 has a very high probability (49$\%$) to assess Company D to category $C_3$ (see Table \ref{Category}), so the DM4 almost agrees with the other DMs. Consequently, Company D has been assigned to category $C_3$.

\begin{table}[htbp]\caption{Results of SMAA-TRI for each DM with $\lambda$ varying in the range $[0.65,0.85]$.}
\centering
   \begin{tabular}{|c||cccccc|}
\hline    
      Company     & DM1 & DM2 & DM3 & DM4 & DM5 & Category Result \\
    \hline
      A & $C_4$(64\%)     & $C_4$(100\%)    & $C_4$(100\%)  &$C_4$(77\%) & $C_4$(73\%)  & $C_4$  \\
      B  & $C_4$(74\%)     & $C_4$(65\%)    & $C_5$  (100\%)  &$C_5$(48\%) & $C_5$(60\%)  & $C_5$  \\
      C & $C_5$(89\%)     & $C_4$(46\%)    & $C_5$(100\%)  &$C_5$(100\%) & $C_3$(69\%)  & $C_5$ \\
      D  & $C_3$(63\%)     & $C_3$(47\%)    & $C_4$(57\%)  &$C_4$(51\%) & $C_3$(55\%)  & $C_3$  \\
    \hline
    \end{tabular}
    \label{Results-Sum-up}%
  \end{table}%

To avoid these conflict situation among the DMs, we consider, as already done in the paper by  \citealp{Morais}, the minimum and the maximum weight for each criterion for all the DMs.

In this way, the weight of each criterion ranges in an interval reflecting all the DMs' attitudes.
The  results obtained are showed in Table \ref{WeightsAllDMs}.

\begin{table}[htbp]\caption{Results of SMAA-TRI by considering  all the DMs.}
\label{WeightsAllDMs}
  \centering
  \resizebox{18cm}{!}{\subtable[Interval weights. ]{\begin{tabular}{|cccccccccccc|}
    \hline
     $g^1_{(1)}$  & $g^1_{(2)}$ & $g^2_{(3)}$ & $g^3_{(4)}$ & $g^3_{(5)}$  & $g^4_{(6)}$ & $g^4_{(7)}$ & $g^5_{(8)}$ & $g^5_{(9)}$ & $g^5_{(10)}$ &  $g^5_{(11)}$ & $g^5_{(12)}$\\  
\hline

      $[0.022,0.112]$ & [0.022,0.112]& [0.0934,0.165]& [0.019,0.174] & [0.019,0.184] & [0.035,0.196] & [0.023,0.181] & [0.023,0.112] &[0.023,0.112] & [0.0632,0.125] & [0.0632,0.125] & [0.0632,0.125] \\ 
    \hline
    \end{tabular}}}\quad
  \subtable[Category acceptability indices.\label{CatAccAllDMs}]{
   \begin{tabular}{|c||cccccc|}
\hline    
   Company & Class 1 & Class 2 & Class 3 & Class 4 & Class 5 & Category Result\\
    \hline
      A & 0\%     & 0\%     & 10\%  & 90\%  & 0\% & $C_4$\\
     B   & 0\% & 6\% & 20\% & 34\%     & 40\% & $C_5$\\ 
     C &   0\%     &  0\%    & 31\%    &  35\%    & 34\% &$C_5$ \\
       D &   0\%     & 0\%     & 40\%     & 50\%     & 10\%  & $C_4$\\
    \hline
    \end{tabular}}%
  \end{table}%

\begin{figure}\caption{Histogram of the category acceptabilities for all the DMs.}
	\centering
		\includegraphics[width=10cm,height=6cm]{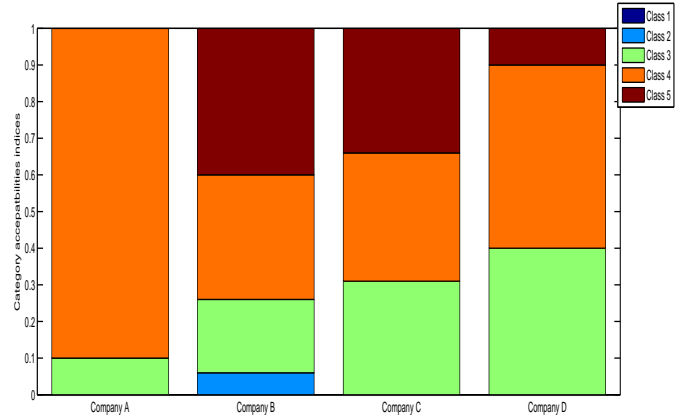}
	\label{histo_AllDMs}
\end{figure}

By performing again the JSMAA software, we have computed the category acceptabilities presented in Table \ref{CatAccAllDMs} (see also Figure \ref{histo_AllDMs} for a representation of category acceptabilities in terms of an histogram).

Looking at Table \ref{CatAccAllDMs}, the assignments of companies A and B are the same given at an individual level (see Table \ref{Results-Sum-up}). Instead, the category results of Companies C and D are not very stable at a group level.
Considering Company C, Table \ref{CatAccAllDMs} shows two similar probabilities of assignment to classes  $C_4$ (35$\%$) and $C_5$ (34$\%$), but the  DMs have decided to assign Company C to class $C_4$ giving more relevance to the individual level.
The same reasoning can be done for Company D. 
From Table \ref{Results-Sum-up}, Company D is  assigned to class 
$C_4$ (50$\%$), even if there is a high probability (40$\%$) which assigns Company D to class $C_3$.

The elaboration of the category acceptabilities and the sorting results have shown some weaknesses of companies C and D. In the following section, we will show how the proposed multicriteria model can be used as a risk assessment tool to reveal such behaviors and detect   the risk factors that can affect the creditworthiness of a company.\\

Furthermore, it is also worthy to notice that  the proposed  credit model can involve some type I and II errors. The  type I and type II errors evaluated in every credit risk model can be interpreted in our model in this way.  The assignment to an ordered class gives a judgment on the creditworthiness of the enterprises.
For example, in our case study classes with the low level of risk are the classes 4 and 5.
 Consequently, if an alternative $i$ is assigned to class 4 or 5, in our model the type I error refers to the classification of a low risk company as a high risk company in all the simulations outperformed. Thus, the error of type I can be evaluated by the sum of the category acceptabilities $\pi^1_i$, $\pi^2_i$ and $\pi^3_i$.
Conversely, if an alternative $i$ is assigned to class 1, 2 or 3, in our model the type II error refers to the classification of high risk firms  as  low risk firms considering all the simulations. Thus, the error of type II can be considered as the sum of the category acceptabilities $\pi^4_i$ and $\pi^5_i$.
For example, if an alternative $i$ has a $\pi^4_i=90\%$ and $\pi^2_i=3\%$ and $\pi^3_i=7\%$, this means that alternative $i$ is sorted to category $C_4$, observing a type I error equal to $10\%$.

\section{Discussion and managerial implications}

On the basis of the results obtained in the previous section, some observations can be done.

From Table \ref{WeightsAllDMs}, one could observe that some inconsistencies could be detected comparing the results obtained from the  aggregation process of all the DMs, based on a common set of intervals of the weights,  to those of the individual DMs shown in Table \ref{Category}.
For instance, is it possible that the aggregate acceptability index for a particular category is positive even if all the corresponding individual indices are zero (and vice versa). As an example take Company D: in Table \ref{WeightsAllDMs}, the acceptability index for class $C_5$ is 10$\%$ even though for 4 of the 5 DMs the indices are zero. On the other hand, take Company B the acceptability index for class $C_2$ is only 6$\%$, but again for 3 of the 5 DMs the indices are $0\%$.
These eventual conflict situations can be explained as follows.
Since for each DM  a different set of weights  has been adopted, in many cases such as in the example of Company B there isn't a consensus of majority of DMs to determine the risk class.



It is worthy to notice that the category acceptability indices could exhibit some "\emph{non-monotonic}" behavior. For instance, take  Company D and suppose to improve the evaluation on the criterion $g^2_{(3)}$ by giving a qualitative score  equal to 4.

In this case, for the DM5 Company D will be assigned either in class $C_3$ or class $C_5$, respectively, with acceptabilities indices of 63$\%$ and 47$\%$. 


In fact, for example take Company D and the DM5, in this case it results that  
$\sigma(D,b_3)= \sigma(D,b_4)=0.666$ and $\sigma(b_3,D)= \sigma(b_4,D)=0.26$. 
From a technical point, such behavior can be explained  by the special features of the ELECTRE methods.
Since Company D overcomes both the profiles $b_3$ and $b_4$, the indices aforementioned are equal.
Then, the assignment of Company D to class $C_3$ or $C_5$ depends on the choice of $\lambda$, expressing the level of credibility with which a company is assigned to a class or to another.
Analyzing the credit granting process, this phenomenon implies that some criteria are the risk factors which
influence the success of a project.
Coming back to the example, the most critical element is the criterion \textit{key competitors} which is at a level 
below the profile $b_3$. 
Even more a company exhibiting a "\emph{non-monotonic}" behavior, it's more fragile from a credit risk point of view since it changes abruptly from a class risk to another not consecutive depending only on the lambda picked at each simulation. In this example, Company D  jumps from class 3 to 5 any time the $\lambda$ considered at each simulation is below $0.778$. Such phenomenon   reveals  a  weakness of this enterprise.

The last example  has suggested us to enrich  the analysis by considering  also imprecise evaluations on the criteria.
In fact, since for innovative projects the  risks  are mostly due to the uncertainties in the data, we have considered  the criteria expressed  in terms of intervals. 
In the case study considered, a few qualitative criteria  have been considered precise especially the ones expressed in a binary code.  For the other qualitative data, we suppose that the evaluations of considered
alternatives on each criterion are integer numbers within an interval. For example, the evaluation of
Company A on criterion $g^2_{(3)}$ can be 3, 4 or 5. For the financial criteria, to be prudential we have lowered their evaluations by 20$\%$ (see Table  \ref{EvIntervals}).
 
\begin{table}[htbp]
  \centering
  \caption{Evaluation matrix with the criteria expressed in terms of intervals.}
  \resizebox{16cm}{!}{  \begin{tabular}{|c||cccccccccccc|}
   \hline
  Company    & $g^1_{(1)}$  & $g^1_{(2)}$ & $g^2_{(3)}$ & $g^3_{(4)}$ & $g^3_{(5)}$  & $g^4_{(6)}$ & $g^4_{(7)}$ & $g^5_{(8)}$ & $g^5_{(9)}$ & $g^5_{(10)}$ &  $g^5_{(11)}$ & $g^5_{(12)}$\\  
    \hline
    \hline
  A&  4     & 3     & $[3,5]$     & $[4,5]$     & $[3,4]$     & 0     & 0     & $[0.44,0.55]$  & $[0.05,0.06]$  & $[0.19,0.24]$  & $[0.18,0.22]$  & $[0.59,0.74]$ \\
  B&  4     & 5     &  $[4,5]$     & $[4,5]$     & $[1,3]$     & 0     & 1     & $[0.58,0.72]$  & $[0.14,0.17]$  & $[0.02,0.03]$  & $[0.12,0.14]$  & $[0.41,0.51]$ \\
  C &  5     & 3     & $[4,5]$     & $[2,4]$     & $[4,5]$     & 1     & 1     & $[0.14,0.18]$  & $[0.04,0.05]$  & $[0.75,0.94]$  & $[0.3,0.36]$   & $[0.45,0.56]$ \\
  D &  4     & 5     & $[3,5]$     & $[2,3]$     & $[4,5]$     & 1     & 0     & $[0.05,0.06]$  & $[0.11,0.14]$  & $[0.42,0.52]$  & $[0.11,0.13]$  & $[0.21,0.26]$ \\
   \hline
   \end{tabular}}%
  \label{EvIntervals}%
\end{table}%
 
 The category acceptabilities obtained are shown in Table \ref{CatAccInterval}.

 \begin{table}[htbp]
 \centering
 \caption{Category acceptabilities with the evaluations criteria in terms of intervals}
\label{CatAccInterval}
\begin{tabular}{|c||cccccc|}
\hline    
   Company & Class 1 & Class 2 & Class 3 & Class 4 & Class 5 & Category Result\\
    \hline
      A & 0\%     & 0\%     & 10\%  & 90\%  & 0\% & $C_4$ \\
     B   & 0\% & 0\% & 10\% & 40\%     & 50\%  & $C_5$\\ 
     C &   0\%     &  0\%    & 17\%    &  50\%    & 34\%  & $C_4$\\
       D &   0\%     & 0\%     & 40\%     & 26\%     & 35\% & $C_3$\\
    \hline
    \end{tabular}
  \end{table}%
Let us observe that depending on the decisional point of view of the bank, the credit officers could fix the lowest probability that allows a category to which a company has to be sorted.
For example, if $\pi^4_i >0.7$ then the company $i$ is assigned to the class $C_4$ (see Table  \ref{CatAccInterval}).
In our case study, this means that only  Company A will be sorted to the class $C_4$.

Moreover, the combination of the imprecision on the criteria and the simulation process  shows to be a useful risk assessment tool since it reveals some weaknesses of  companies C and D.
  
In fact, while the assignments of companies A and B are unchanged, the assignments of Company C and especially Company D are changed significantly (see Table  \ref{CatAccInterval}). 
The uncertainty on the data has lowered the categories of companies C and D  comparing with the category results at a group level (see Table \ref{CatAccAllDMs}). Such results can be linked to the the remark on the non-monotonic behavior considered above.

At this point,  it could be useful an analysis of the mortality rates of the enterprises of the sectors considered to 
strengthen or lower the evaluation given by the multicriteria model.
At this aim, we have reported in Table \ref{Mortality} the mortality rates for the sectors under consideration over the period 2008-2012 and we have compared them to the ones of all the Italian SMEs.
 \begin{table}[htbp]
 \centering
 \caption{Mortality rates (data from ISTAT)}
\label{Mortality}
\begin{tabular}{|c||ccccc|}
\hline    
     & 2008 & 2009 & 2010 & 2011 & 2012 \\
    \hline
   Sector Company    A   & 5.2\% & 4.9\%     & 4.9\%     & 4.9\% & 4.8\%  \\
    Sector Company   B   & 9.1\% & 7.5\% & 7.8\% & 8.1\% & 7.5\%\\ 
    Sector Company   C   & 12.8\% &   9.6\%     &  11.4\%    & 9.9\% & 9.2\%   \\
    Sector Company   D   & 10.8\% &   9.6\%     & 11.4\%     & 10.7\% & 11.3\%   \\
    \hline
    \hline
       Italian SMEs      & 7.1\% & 7.9\%  & 7.8\% & 7.7\% & 8.1\% \\
      \hline
    \end{tabular}
  \end{table}%
 
From a macroeconomic point of view  also this analysis reveals a high risk to finance  enterprises C and D since the mortality rates of C and D sectors are greater than those at a national level.

Finally, it is still important to emphasize that in the judgmental credit risk model considered, 
we have never introduced any veto criterion. The introduction of veto criteria is expected to have a considerable impact on the decision procedure. In fact, let us suppose for example to consider the criterion  availability of testing and unit pilots ($g^4_{(6)}$) as a veto criterion. 
One implication of the inclusion of this veto criterion is that, when an enterprise doesn't have a unit pilot (thus receiving a $0$ point in the evaluation according to this criterion), the criterion would be against the assignment of such an enterprise to a good risk class. Such enterprise would be assigned to the class $C_1$,  even if it has a good evaluation on the majority of criteria.
In the case study, for example only the enterprise Company C has a good evaluation on this criterion.
As a result, in many real situations, it may be very relevant to impose some veto criteria.

\section{Conclusions}
Many recent studies have been dedicated to measuring SMEs' innovation considered as a crucial factor for the development of a national economy (\citealp{BankItaly}).
To achieve a high level of innovation the main obstacle  that the innovative SMEs face is a non straightforward access to credit.
For example in Italy, such asymmetric information is due to the small  dimension of most of the innovative SMEs (\citealp{Cannari}).
To help innovative enterprises, financial institutions could try to reduce such informative asymmetries.
Presently, banks assess their credit risk by evaluating the business plan and considering some non-financial information such as  the market trend and the quality of the management team.
Although financing innovation isn't an easy task due to the lack of track records of innovative SMEs, we believe that a  rating model based upon experts' judgments could improve it. In financing innovative SMEs, the role of the experts is crucial, since they can help the credit officers in selecting the proper criteria, especially, in detecting their risks.

In this paper, we have presented a multicriteria approach to sort innovative enterprises into risk classes, mainly on the basis of soft information.

Specifically, we describe a possible hierarchical structure of the innovation risks: development, production, market and technological ones.
Such risk indicators have been considered together with some financial criteria.
The multicriteria approach proposed is ELECTRE-TRI based on an outranking preference relation comparing each innovation to some existing risk profiles. As explained in the paper, the multidimensional and complex decisional framework  of financing innovation is well adapted to the  aforementioned multicriteria method.

In fact, the credit officers with the help of the experts could define the risk profiles, the criteria weights, but could also  tune some specific parameters such as the cutting level, the preference, indifference and veto thresholds on which the final ratings depend.

Finally, we envisage some possible  research lines for the future:
\begin{itemize}
\item It could be interesting to detect the most ``critical'' criteria governing the decision making process.
What is meant 
by the term ``critical'' here is the smallest change that 
might occur to a certain criterion
in order increase  the category acceptability of the better risk class to which each enterprise has been assigned.
  \item  ELECTRE-TRI is a non-compensatory method; maybe it could be useful to consider  also other multicriteria models including interaction between criteria (see the Choquet integral preference model among the multicriteria approaches representing interaction between criteria e.g. in \citealp{angilella2004assessing} and its recent extension within a SMAA methodology in \citealp{Angilella2012}).
  \item Since the criteria adopted in this paper are hierarchally structured, it could be useful to apply the recent approach of the hierarchal Choquet integral  presented in \citealp{acg2013}.

\end{itemize}

\section*{Acknowledgments}
The authors would like to thank some anonymous  reviewers for their comments and suggestions which yielded a considerable improvement of  the manuscript.
The authors thank  Professor A. Sironi for some useful discussions and suggestions on the paper.   
%

%
%

%



\bibliographystyle{plainnat}

\bibliography{Creditbiblio}
\newpage

\section*{Appendix A }\label{appendixA}

%
\begin{table}[htbp]
  \centering
  \caption{Scenario analysis\label{Unica}}
    \resizebox{8cm}{!}{\subtable[Company A \label{ScenarioBIOSOIL}]{ \begin{tabular}{|c||c|c|c|c|c|}  
    \hline
\multicolumn{6}{|c|}{\textbf{Initial cash flows}}\\ 
   \hline
   \hline
    & year 1 & year 2 & year 3 & year 4 & NPV ($7.93\%$) \\
    \hline
    \hline
   	   cash-flow (\euro)  & -43,534.00	&   69,616.91 	& 	 9,178.96 &		 118,470.63	& 140,275.51 \\
  \hline
  \multicolumn{6}{|c|}{\textbf{Cash flows lowered by 20$\%$}}\\
\hline
\hline
& year 1 & year 2 & year 3 & year 4 & NPV ($7.93\%$) \\
\hline
  \hline
    cash-flow (\euro) & -52,240.80	& 55,693.53	& 7,343.17	& 94,776.50	&  € 75,092.75\\
\hline
  \multicolumn{6}{|c|}{\textbf{Cash flows lowered by 40$\%$}}\\ 
  \hline
  \hline
  & year 1 & year 2 & year 3 & year 4  & NPV ($7.93\%$) \\
    \hline
  \hline
  cash-flow (\euro) & -60,947.60	& 41,770.15	& 5,507.37	& 71,082.38	& 36,151.86\\
\hline
  \end{tabular}}
  }\resizebox{10cm}{!}{\subtable[Company B \label{ScenarioDowSee}]{
    \begin{tabular}{|c||c|c|c|c|c|c|}  
    \hline
\multicolumn{7}{|c|}{\textbf{Initial cash-flows}}\\ 
   \hline
   \hline
    & year 1 & year 2 & year 3 & year 4 & year 5 & NPV ($7.93\%$) \\
    \hline
    \hline
   cash-flow (\euro)  & - 8,715.00	&  -15,528.00 	& 	52,196.00 &	58,422.00		& 57,472.00 & 102,405.74\\
  \hline
  \multicolumn{7}{|c|}{\textbf{Cash flows lowered by 20$\%$}}\\
\hline
\hline
& year 1 & year 2 & year 3 & year 4 & year 5 & NPV ($7.93\%$) \\
  \hline
  \hline
 cash-flow (\euro) & - 10,458.00 &	- 18,633.60	& 41,756.80	& 46,737.60 & 45,977.60 &73,362.71\\
\hline
  \multicolumn{7}{|c|}{\textbf{Cash flows lowered by 40$\%$}}\\ 
  \hline
  \hline
& year 1 & year 2 & year 3 & year 4 & year 5 & NPV ($7.93\%$) \\
  \hline
  \hline
  cash-flow (\euro) & - 12,201.00	& - 21,739.20	& 31,317.60	& 35,053.20	 & 34,483.20 & 44,319.68\\
\hline
  \end{tabular}}
  }
  
  \resizebox{10cm}{!}{\subtable[Company C \label{ScenarioMRS}]
{ \begin{tabular}{|c||c|c|c|c|c|c|c|} 
   \hline
   \multicolumn{8}{|c|}{\textbf{Initial cash flows}}\\ 
   \hline
   \hline
    & year 1 & year 2 & year 3 & year 4 & year 5 & year 6 &  NPV ($7.93\%$) \\
    \hline
    \hline
      cash-flow (\euro) & -211,100.00	 & 126,543.19 	 & 196,034.36 	 & 233,763.68 	 & 438,942.51 	 & 568,339.16 &  1,032,614.23
\\
\hline
  \multicolumn{8}{|c|}{\textbf{Cash flows lowered by 20$\%$}}\\
\hline
\hline
  & year 1 & year 2 & year 3 & year 4 & year 5 & year 6 &  NPV ($7.93\%$) \\
    \hline
    \hline
   cash-flow (\euro) &  -253,320.00 &	101,234.55 &	156,827.48 &	187,010.95	& 351,154.01 &	454,671.33 & 642,152.30\\
  \hline
  \multicolumn{8}{|c|}{\textbf{Cash flows lowered by 40$\%$}}\\ 
  \hline
  \hline
  & year 1 & year 2 & year 3 & year 4 & year 5 & year 6 &  NPV ($7.93\%$) \\
    \hline
    \hline
   
  cash-flow (\euro) &   -295,540.00	 & 75,925.91	& 117,620.61	& 140,258.21 & 	263,365.51 & 	341,003.49 & 383,819.35\\
\hline
  \end{tabular}}
  } \resizebox{8cm}{!}{\subtable[Company D \label{ScenarioNANO}]{  \begin{tabular}{|c||c|c|c|c|}  
    \hline
   \multicolumn{5}{|c|}{\textbf{Initial cash-flows}}\\ 
   \hline
   \hline
   & year 1 & year 2 & year 3 & NPV ($7.93\%$) \\
    \hline
    \hline
  cash-flow (\euro) & -62,272.07	& 204,057.11 &	1,094,740.87 & 988,208.95\\

  \hline
  \multicolumn{5}{|c|}{\textbf{Cash flows lowered by 20$\%$}}\\
\hline
\hline
& year 1 & year 2 & year 3 & NPV ($7.93\%$) \\
    \hline
    \hline
  cash-flow (\euro) & -74,726.48	& 163,245.68	& 875,792.69 & 767,488.48\\
\hline
\hline
  \multicolumn{5}{|c|}{\textbf{Cash flows lowered by 40$\%$}}\\

  \hline
  \hline
    & year 1 & year 2 & year 3 & NPV ($7.93\%$) \\
  \hline
  \hline

cash-flow (\euro) & - 87,180.89 & 122,434.26  & 656,844.52 & 546,768.00\\
\hline
  \end{tabular}}
  }%
  
  \end{table}%
	
	\begin{table}[!h]
\centering
\caption{Comparisons with the sector.  \label{financialindicators}}
\resizebox{12cm}{!}{\subtable[Company A\label{BIOSOILSector}]{
    \begin{tabular}{|c|r|r|r|r|r|}
    \hline
    & Intangible Asset/Fixed Asset & R$\&$D/Sales & ROA   & Short term debt/Equity  &  Cash/Total Asset \\
    \hline
     Company A& {$0.55$} & {$0.06$}& {$0.24$} &  {$0.18$} &  {$0.74$}  \\
    \hline
    \hline
    \textbf{ min} & $0.00$ & -& $-0.03$ &  $0.14$ &  $0.02$  \\
    \hline
    \textbf{Quartile 25} & $             0.02 $ & - & 0& $-        0.03 $  & $ 0.14 $   \\
    \hline
    \textbf{Quartile 50} & $0.01 $ & - & $          0.01 $ & $            1.42 $ &             $  0.07 $  \\
    \hline
    \textbf{Quartile 75} & $0.20 $ & -& $        0.05 $ &                      5.44  &               $0.18 $  \\
    \hline
    \textbf{max} & $0.20$ & -  & $0.05$ & $5.44$ &  $0.18$  \\
    \hline
    \end{tabular}}%
    }
 \resizebox{12cm}{!}{\subtable[Company B\label{DowSeeSector}]{
 \begin{tabular}{|c|r|r|r|r|r|}
    \hline
    & Intangible Asset/Fixed Asset & R$\&$D/Sales & ROA &   Short term debt/Equity  &  Cash/Total Asset \\
    \hline
  Company B & $0.72$ & $0.17$ & $0.03$ &  $0.12$ &  $0.51$ \\
  \hline
  \hline
    { min} & {$0.00$} & -& {$-0.01$} &  {$0.31$} &  {$0.02$}  \\
    \hline
    {Quartile 25} & {$0.00$} & -& {$-0.01$} &  {$0.31$} &  {$0.02$}  \\
    \hline
    {Quartile 50} & {$0.17$} & {-}& {$0.03$} &  {$1.22$} &  {$0.07$}  \\
    \hline
    {Quartile 75} & {$1.34$} & {-  }& {$0.09$} &  {$3.72$} &  {$0.18$}  \\
    \hline
    { max}& {$1.34$} & {-}  & {$0.09$} &  {$3.72$} &  {$0.18$} \\
     \hline
    \end{tabular}
     }}
 \begin{center}
\resizebox{12cm}{!}{\subtable[Company C\label{MRSSector}]{
    \begin{tabular}{|c|r|r|r|r|r|}
    \hline
    & Intangible Asset/Fixed Asset & R$\&$D/Sales & ROA   & Short term debt/Equity  &  Cash/Total Asset \\
    
    \hline
   Company C & $0.18$ & $0.05$ &  {$0.94$} &  $0.3$ &  $0.56$  \\
    \hline
    \hline
    {min} & {$0.00$} & -& {$-0.01$} & {$0.28$} &  {$0.01$}  \\
    \hline
    {Quartile 25} & {$0.00$} & -& {$-0.01$} & {$0.28$} &  {$0.01$}  \\
    \hline
    {Quartile 50} & {$0.09$} & {-} & {$0.04$} &  {$1.07$} & {$0.06$} \\
    \hline
    {Quartile 75} & {$0.43$} & {-} & {$0.10$} &  {$3.14$ } &  {$0.16$} \\
    \hline
    { max} & {$0.43$} & {-} & {$0.10$} &  {$3.14$} &  {$0.16$} \\
          
    \hline
    \end{tabular}%
 }}
\resizebox{12cm}{!}{ \subtable[Company D\label{NanoTechSector}]{
\begin{tabular}{|c|r|r|r|r|r|}
    \hline
    & Intangible Asset/Fixed Asset & R$\&$D/Sales & ROA   & Short term debt/Equity  &  Cash/Total Asset \\
     \hline
         Company D & 0.06 & 0.14& {0.52} & 0.11 & 0.26 \\
          \hline
    \hline
    { min} & {$0.00$} & -& {$-0.04$} &   {$0.14$} &  {$0.03$}  \\
    \hline
    {Quartile 25} & {$0.00$} & - & {$-0.04$}   & {$0.14$} &  {$0.03$} \\
    \hline
    {Quartile 50} & {$0.07$} & {-} & {$0.00$} & {$0.67$}  & {$0.08$} \\
    \hline
    {Quartile 75} & {$0.91$} & {-}& {$0.04$} &  {$2.55$} &  {$0.21$}  \\
    \hline
    {max} & {$0.91$} & {-} & {$0.04$} &  {$2.55$} &  {$0.21$} \\
    \hline
    \end{tabular}%
  }
 }\end{center}
\end{table}

\newpage
\section*{Appendix B }\label{appendixB}

To assess the criteria weights within a Simos'procedure, the following variables are defined:
\begin{itemize}

\item $z$ is the ratio expressing how many times the last criterion is more important than the first one in the ranking;

\item $e'_r$ is the number of white cards between the rank $r$ and $r+1$;

\item $e_r=e'_r+1$;

\item $u=\frac{z-1}{e}$;

\item $\displaystyle e=\sum_{r=1}^{n-1}e_r$.
\end{itemize}

The non normalized weight $k(r)$ is computed by:
$$k(r)=1+u(e_0+\cdots+e_{r-1}),$$ 
with $e_0=0.$

Let $k'_i=k(r)$ be the weight relative to the criterion $i$ and let $\displaystyle K'=\sum_{i=1}^nk_i'$ the sum of the non-normalized weights. The normalized weight $k_i^*$ is computed by:

$$k^*_i=\frac{1}{K'}k'_i.$$

Within the Simos' procedure, the preference information expressed by the DM1 is reported in Table \ref{simosappendixone}. 
, the normalized weights are  those displayed in Table \ref{simosappendixtwo}.

\begin{table}[htbp]
  \caption{Simos' method}
 \centering
\subtable[DM1's set of weights with $z=8$.\label{simosappendixone}]{
    \begin{tabular}{|cccccc|} 
    \hline
    Rank r & Criteria  & n. of white & $e_r$  & Non-normalized   & Total \\
       &     in the rank $r$       &   cards                &          & weights $k(r)$ & \\
       \hline
       
    1     & $\{g^1_{(1)},g^1_{(2)}\}$    & 0     & 1     & 1     & 2 \\
    2     & $\{g^5_{(8)},g^5_{(9)}\}$ & 0     & 1     & 1.63 & 3.27 \\
    3     & $\{g^3_{(4)},g^3_{(5)}\}$   & 0     & 1     & 2.27 & 4.54 \\
    4     & $\{g^5_{(10)},g^5_{(11)}, g^5_{(12)} \}$ & 5     & 6     & 2.90 & 8.72 \\
    5     & $\{g^2_{(3)}\}$     & 0     & 1     & 6.72 & 6.72 \\
    6     & $\{g^4_{(7)} \}$      & 0     & 1     & 7.36 & 7.36 \\
    7     & $\{g^4_{(6)} \}$     &       &       & 8     & 8 \\    
  \hline
Total        &       &     $e$  &     11  &   $K'$   & 40.63 \\
    \hline
    \end{tabular}}\quad\quad
\subtable[Normalized weights for the DM1.\label{simosappendixtwo}]{%
    \resizebox{16cm}{!}{\begin{tabular}{|r||rrrrrrrrrrrr|}
    \hline
   & $g^1_{(1)}$  & $g^1_{(2)}$ & $g^2_{(3)}$ & $g^3_{(4)}$ & $g^3_{(5)}$  & $g^4_{(6)}$ & $g^4_{(7)}$ & $g^5_{(8)}$ & $g^5_{(9)}$ & $g^5_{(10)}$ &  $g^5_{(11)}$ & $g^5_{(12)}$\\  
\hline
$k^*_i$ &   0.0250 & 0.0250 & 0.1650 & 0.0560 & 0.0560 & 0.1960 & 0.1810 & 0.0400 & 0.0400 & 0.0720 & 0.0720 & 0.0720 \\
    \hline
    \end{tabular}}}%
  \end{table}%

\end{document}